\documentclass[conference]{IEEEtran}
\IEEEoverridecommandlockouts
\usepackage{cite}
\usepackage{amsmath,amssymb,amsfonts}
\usepackage{algorithmic}
\usepackage{graphicx}
\usepackage{textcomp}
\usepackage{xcolor}
\usepackage{multirow}

\def\BibTeX{{\rm B\kern-.05em{\sc i\kern-.025em b}\kern-.08em
    T\kern-.1667em\lower.7ex\hbox{E}\kern-.125emX}}

\usepackage{tikz}
\usepackage{textcomp}
\usepackage[doipre={DOI:~}]{uri}
\usepackage{lipsum}
\newcommand\copyrighttext{%
  \footnotesize \textcopyright 2024 IEEE. Personal use of this material is permitted.  Permission from IEEE must be obtained for all other uses, in any current or future media, including reprinting/republishing this material for advertising or promotional purposes, creating new collective works, for resale or redistribution to servers or lists, or reuse of any copyrighted component of this work in other works.}
\newcommand{\copyrightnotice}{%
\begin{tikzpicture}[remember picture,overlay]
\node[anchor=south,yshift=10pt] at (current page.south) {\fbox{\parbox{\dimexpr\textwidth-\fboxsep-\fboxrule\relax}{\copyrighttext}}};
\end{tikzpicture}%
}

\definecolor{somegray}{rgb}{0.5, 0.5, 0.5}
\newcommand{\darkgrayed}[1]{\textcolor{somegray}{#1}}
\makeatletter
\newcommand*\titleheader[1]{\gdef\@titleheader{#1}}
\AtBeginDocument{%
  \let\st@red@title\@title
  \def\@title{%
    \vskip-2.0em
    \bgroup\normalfont\large\centering\@titleheader\par\egroup
    \vskip0.0em\st@red@title}
}
\makeatother

\titleheader{\darkgrayed{This paper has been accepted for publication in the AICAS 2025 conference. \copyright 2024 IEEE}}

\title{EnhancePPG: Improving PPG-based Heart Rate Estimation with Self-Supervision and Augmentation}

\author{\IEEEauthorblockN{
    Luca Benfenati\IEEEauthorrefmark{1},
    Sofia Belloni\IEEEauthorrefmark{1},
    Alessio Burrello\IEEEauthorrefmark{1}\IEEEauthorrefmark{5},
    Panagiotis Kasnesis\IEEEauthorrefmark{2}\IEEEauthorrefmark{3},
    Xiaying Wang\IEEEauthorrefmark{4},\\
    Luca Benini\IEEEauthorrefmark{4}\IEEEauthorrefmark{5},
    Massimo Poncino\IEEEauthorrefmark{1},
    Enrico Macii\IEEEauthorrefmark{6},
    Daniele Jahier Pagliari\IEEEauthorrefmark{1}}

    \vspace{0.2cm}
    
    \IEEEauthorblockA{\IEEEauthorrefmark{1}DAUIN, Politecnico di Torino, Italy, \IEEEauthorrefmark{2}DEEE, University of West Attica, Egaleo, Greece,\\ \IEEEauthorrefmark{3}ThinGenious PC, Marousi, Greece,\IEEEauthorrefmark{4}IIS, ETH Z{\"u}rich, Z{\"u}rich, Switzerland,\\ \IEEEauthorrefmark{5}DEI, University of Bologna, Bologna, Italy, \IEEEauthorrefmark{6}DIST, Politecnico di Torino, Italy}

    \vspace{-0.8cm}
    }

\begin{document}
\maketitle
\copyrightnotice
\begin{abstract}
Heart rate (HR) estimation from photoplethysmography (PPG) signals is a key feature of modern wearable devices for health and wellness monitoring. While deep learning models show promise, their performance relies on the availability of large datasets. We present EnhancePPG, a method that enhances state-of-the-art models by integrating self-supervised learning with data augmentation (DA). Our approach combines self-supervised pre-training with DA, allowing the model to learn more generalizable features, without needing more labelled data. Inspired by a U-Net-like autoencoder architecture, we utilize unsupervised PPG signal reconstruction, taking advantage of large amounts of unlabeled data during the pre-training phase combined with data augmentation, to improve state-of-the-art models' performance.
Thanks to our approach and minimal modification to the state-of-the-art model, we improve the best HR estimation by 12.2\%, lowering from 4.03 Beats-Per-Minute (BPM) to 3.54 BPM the error on PPG-DaLiA. Importantly, our EnhancePPG approach focuses exclusively on the training of the selected deep learning model, without significantly increasing its inference latency.  
\end{abstract}


\section{Introduction \& Related Works}\label{ch:introduction}
Wearable health monitoring devices have rapidly advanced in the health and wellness Internet-of-Things (IoT) era, integrating sophisticated sensors, efficient computational platforms, and cutting-edge algorithms \cite{iot1, iot2, iot3}. Heart Rate (HR) is a vital physiological parameter that offers insights into physical activity, stress levels, and potential health concerns \cite{health-random}. Although HR can be assessed intermittently in clinical or home environments, continuous monitoring is essential for the early detection of health issues.

One way to continuously monitor HR is wrist-worn Photoplethysmographic (PPG) sensors. These sensors have become popular with the rise of affordable fitness trackers like Apple Watch \cite{applewatch} and Fitbit \cite{fitbit}. They measure HR by detecting light from pulsating light-emitting diodes (LEDs) absorbed or reflected by blood vessels \cite{ppg-review}. Despite their user comfort, PPG sensors face accuracy challenges due to Motion Artifacts (MA) from variations in sensor pressure, external light interference, and arm movements, particularly during intense physical activity when blood flow variability complicates HR estimation.

To address the challenges posed by MA, initial approaches in the literature proposed filtering techniques, where the correlation between acceleration data and PPG signals was leveraged to mitigate MA noise on PPG signal\cite{robustPPG}, \cite{robustPPG2}. However, these methods often suffer from poor generalization on unseen data due to the high number of tunable hyperparameters.

Deep learning techniques have been proposed to overcome this limitation, showing promising results on various public datasets \cite{ppg_dalia}. In \cite{qppg, embedPPG}, the authors address the challenges of poor generalization and the high complexity of deep learning by integrating Neural Architecture Search (NAS) strategies with Temporal Convolutional Networks (TCNs) for HR monitoring. Despite the progress made, most deep learning solutions fail to achieve a sufficiently low Mean Absolute Error (MAE) in HR estimation for practical deployment in real-world scenarios, where inter-subject variability makes accurate HR monitoring particularly difficult.

In this work, we propose an innovative approach that integrates self-supervised learning and data augmentation techniques \cite{augmentPPG, kid-ppg} to enhance HR estimation performance of deep learning models. While our method is mostly orthogonal to the specific deep neural network selected, we assess its performance using PULSE (\textbf{P}pg and im\textbf{U} signa\textbf{L} fu\textbf{S}ion for HR \textbf{E}stimation \cite{pulse}), the state-of-the-art deep learning model for HR estimation from PPG signals on PPG-DaLiA \cite{ppg_dalia}, the biggest public dataset available. By introducing a pre-training phase combined with augmentation techniques, we improve the robustness and the accuracy of PULSE without significantly impacting inference latency, ensuring suitability for deployment on low-power edge devices like smartwatches. 

Our contributions are as follows:
\begin{itemize}
    \item \textbf{Self-supervised learning (SSL) for PPG data}: we are the first to introduce a self-supervised pre-training step by restructuring the selected deep learning model into an autoencoder-like framework, inspired by U-Net \cite{u-net}. This allowed us to pre-train the model on two additional unlabelled datasets: WESAD \cite{wesad} and a new dataset provided by West Attica University. The pre-training task is the PPG signal reconstruction. This approach allows for the use of more readily accessible data, as they can be collected without needing any labelling.
    \item \textbf{Data Augmentation (DA) Integration}: we leverage DA techniques, inspired by \cite{augmentPPG}, to expand the pre-training dataset with a focus on increasing HR variability, thereby enhancing the model's generalization and robustness to noisy inputs. Additionally, we evaluate how our SSL framework supports the use of DA without altering the HR labels, ensuring that the model can learn broader features without compromising its ability to generalize to unseen data.
\end{itemize}
Ours is the first attempt to exploit large unlabelled datasets combined with data augmentation in a self-supervised way for HR estimation, paving the way to utilising large unlabelled datasets to improve the generalization of this vital task. 
Importantly, our EnhancePPG approach, which combines self-supervised pre-training with data augmentation, solely affects the training phase, leaving the inference process unchanged. To benchmark our method's accuracy, we used a modified PULSE architecture derived from the original one to reduce training time and improve performance. These architectural changes alone lowered the Mean Absolute Error (MAE) from 4.03 BPM to 3.73 BPM. Applying EnhancePPG, we brought the MAE of the architecture down to 3.54 BPM, achieving a further  5.09\% reduction. Notably, we achieved significant reductions in MAE for the two most critical patients, S5 and S8, reducing the MAE achieved by the original PULSE model by 18.9\% and 36.8\%, respectively. 

Our modified PULSE model can be deployed on an STM32 NUCLEO-H743ZI2 with a latency of 431.6 ms, only 4.43\% higher than the original one. On the other hand, enhancing the model with our proposed pipeline does not lead to any latency increment.
\section{Material \& Methods}\label{ch:methods}
\subsection{Model architecture}\label{sec:pulse_arc}
We evaluate our pre-training and data augmentation methodology for PPG-based HR estimation on an enhanced version of PULSE, state-of-the-art model for PPG-based HR estimation originally introduced by \cite{pulse}. PULSE leverages PPG signals and tri-axial accelerometer data to improve HR estimation accuracy by integrating temporal convolutions with a Multi-Head Cross-Attention (MHCA) module. 
The architecture comprises three 1D convolutional blocks with increasing channels (32, 48, 64), each containing three consecutive convolutional layers to capture short-term temporal variations in both PPG and accelerometer data. Each block's output is downsampled using average pooling layers. The feature maps generated by the convolutional layers are processed by the MHCA module, where PPG signals serve as query vectors and accelerometer features are used as key and value vectors. Finally, the attention module's output is normalized and passed through two dense layers to produce the HR estimate. 

\begin{figure}
    \centering
    \includegraphics[width=\columnwidth]{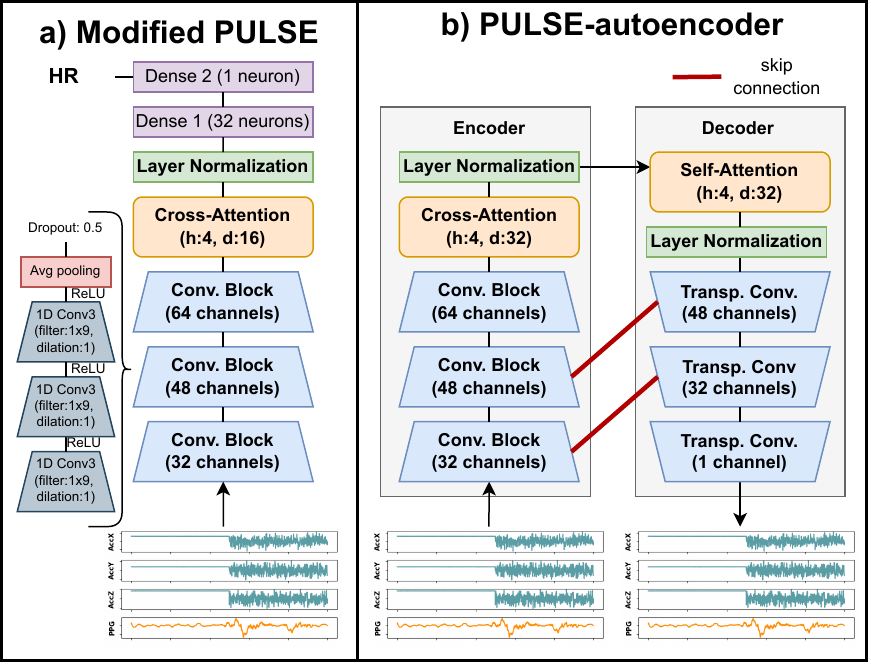}
    \vspace{-0.6cm}
    \caption{a) Modified PULSE architecture used in fine-tuning; b) our PULSE-autoencoder architecture used in pre-training}
    \label{fig:architecture}
    \vspace{-0.5cm}
\end{figure}

We modify the original PULSE architecture by replacing dilated convolutions with standard convolutions. To maintain the same receptive field, we adjust the convolutional kernel size from (1, 5) with a dilation of 2 to (1, 9) with a dilation of 1. This modification (Fig. \ref{fig:architecture}a) allows the model to concentrate on more localized features maintaining a comparable inference latency.

\subsection{Self-supervised pre-training}\label{sec:pretraining}
We employ SSL to leverage the vast quantities of unlabeled PPG data available. In SSL, initially, the model undergoes a pre-training phase, during which it learns the underlying structure of data and extracts task-agnostic feature representations. This pre-trained model can subsequently be fine-tuned for a range of downstream tasks. In our approach, pre-training is essential for improving HR estimation accuracy, especially in situations where motion artefacts could undermine model robustness.
For our SSL pre-training setup, we restructured the PULSE architecture into an autoencoder, compressing and reconstructing input signals, thus forcing the model to capture meaningful latent representations \cite{biseizure}.
The encoder part is based on the PULSE model described in \ref{sec:pulse_arc}, with the two last dense layers, i.e. the regression head, removed.
The decoder mirrors the encoder's structure, using transposed convolutional layers to reconstruct the input signal from the latent space. Similar to the encoder, the decoder utilizes multi-headed attention mechanisms, but in this case, the input tensor is self-correlated. Inspired by U-Net \cite{u-net}, we added two skip connections between the first and the second convolutional blocks of the encoder and decoder to preserve spatial information during up-sampling. 
As a result, the first two convolutional blocks of the decoder increase by a factor 2 both the spatial dimensions and the channels, as they concatenate their own output with the corresponding feature map from the contracting path.
This architecture is shown in Fig. \ref{fig:architecture}b.

During pre-training, the model is optimized in a completely self-supervised way to perform a reconstruction of the input signal, with the Mean Squared Error (MSE) as the objective metric, defined as the average squared difference between the input values and the reconstructed values. The MSE is averaged across the four sensor modalities considered: PPG and acceleration along the $x$, $y$, and $z$ axes. As for the training details, we pre-train our PULSE-autoencoder model for 500 epochs, with AdamW optimizer with a learning rate of \(1e^{-3}\), betas $(0.9, 0.95)$, and a weight decay of $0.01$. We also adopt a learning rate scheduler based on a half-cycle cosine decay when the validation loss fails to improve for $5$ consecutive epochs. Lastly, we use early stopping with $50$ epochs patience.

\subsection{Fine-tuning}\label{sec:finetuning}
Following pre-training, the model is adjusted for fine-tuning on the task-specific dataset by restoring the modified PULSE architecture described in Sec. \ref{sec:pulse_arc}. The weights learned during pre-training for the encoder are used to initialize the HR estimation model. During fine-tuning, the model is trained to minimize the MAE between the predicted HR and the true HR value from the dataset. Model validation follows the Leave-One-Session-Out (LOSO) cross-validation protocol already implemented in \cite{pulse, loso1, loso2}, where data from the subjects are split into four folds. Three folds are used for training, while the remaining one is divided subject-wise, into test and validation sets. This training protocol ensures the generalizability of the produced model to unseen subject data. We fine-tune our model for $500$ epochs, using Adam with a learning rate of $5e^{-4}$ and early stopping with patience of $150$.

\subsection{Augmentation techniques}\label{sec:aug}
We leverage DA techniques to boost our model's performance and robustness. DA helps mitigate overfitting and enhances generalization, which is crucial given the challenges of collecting large, high-quality PPG datasets that require costly medical-grade electrocardiogram (ECG) devices for accurate ground-truth HR measurement.

Augmented data points, $\hat{x}_i$, are derived from the original data $x_i$ through noise injection or transformations like scaling, reshaping, and temporal adjustments. Inspired by \cite{augmentPPG}, we selected their argumentation setup that offers the best performance on critical patients, i.e., the ones presenting the highest HR estimation error due to having an average BPM on the upper or lower limits of the dataset. This configuration employs two PPG-specific DAs: 
\begin{itemize}
    \item the \textit{Divide\_2}, which extends the lower bound of the BPM range by randomly selecting, for each input window of length $T$, a continuous section of length $T /2$, and stretching it in time back to the original window size ($0$ to $T$) with resampling;
    \item the \textit{Multiply\_[x-y, z]}, which extend the upper bound of the BPM range, by multiplying the HR by factors between $x$ and $y$, with a step of $z$ and then undersampling it back to its original window dimension.
\end{itemize}
We specifically evaluated the DA setup comprising \textit{Divide\_2} combined with \textit{Multiply}\_$[1.2-2, 0.1]$, thus increasing the dataset by a factor of 11x. Unlike \cite{augmentPPG}, we apply  DA on pre-training data, avoiding the need to alter HR labels and thereby reducing the risk of inaccuracies and inconsistencies that could degrade model performance. Notably, our work is the first to combine SSL with DA for PPG-based HR estimation, expanding the pre-training dataset to enhance pattern recognition and improve generalization.

\subsection{Datasets}  
Three datasets were used to train and test the models: PPG-DaLiA \cite{ppg_dalia}, WESAD \cite{wesad}, and a newly collected dataset by West Attica University. Signals were segmented into 8-second windows with a 2-second shift.  

\textbf{PPG-DaLiA} includes data from 15 subjects (8 female, 7 male, aged 21–55) performing 8 daily activities (e.g., walking, cycling). Recorded with the Empatica E4 wristband \cite{empatica}, it provides one-channel PPG (64Hz) and tri-axial accelerometer (32Hz) data. Ground truth HR was obtained from a chest-worn smart belt, with ECG-based HR, as the average instantaneous HR within each 8-second window. The dataset contains ~64,697 samples from two hours of recordings per subject.  

\textbf{WESAD} is a multimodal dataset for stress and affect detection, with data from 15 participants (aged 24–35). It includes signals like PPG, ECG, and accelerometer, capturing neutral, stress, and amusement states. Importantly, the dataset labels represent discrete affective states, not HR values, making it unsuitable for direct supervised training in HR estimation tasks. However, through our self-supervised pre-training, these data can be effectively integrated. The dataset provides ~43,385 samples from ~100 minutes of recordings per participant.  

The \textbf{West Attica dataset} contains PPG and accelerometer data from healthy subjects recorded during rest using the Empatica E4. It includes 449,544 samples—10 times larger than the public datasets—without labels, enabling simpler and more scalable data collection for unsupervised learning.

\section{Results}\label{ch:results}
\begin{table*}
\centering
\caption{Per subject MAE [BPM] performance of our approach on PPG-DaLiA compared with state-of-the-art.}
\label{tab:results_soa}
\resizebox{0.95\textwidth}{!}{ 
\begin{tabular}{l|ccccccccccccccc|c}
\hline
\textbf{Model} & \textbf{S1} & \textbf{S2} & \textbf{S3} & \textbf{S4} & \textbf{S5} & \textbf{S6} & \textbf{S7} & \textbf{S8} & \textbf{S9} & \textbf{S10} & \textbf{S11} & \textbf{S12} & \textbf{S13} & \textbf{S14} & \textbf{S15} & \textbf{Mean} \\ 
\hline
AugmentPPG \cite{augmentPPG} & 4.37 & 3.74 & 2.43 & 5.49 & 9.41 & 3.63 & 2.23 & 7.86 & 8.94 & 3.32 & 5.34 & 7.64 & 2.03 & 2.94 & 3.58 & 4.86 \\
PULSE\cite{pulse} & 3.78 & \textbf{3.04} & 2.20 & 4.41 & 6.95 & 3.71 & 2.39 & 8.17 & \textbf{6.19} & \textbf{2.60} & 3.85 & \textbf{5.22} & 1.98 & 3.13 & 2.79 & 4.03 \\
KID-PPG* \cite{kid-ppg} & 4.27 & 3.46 & 2.07 & 5.61 & 3.01 & 2.74 & 1.39 & 7.13 & 9.53 & 2.77 & 3.58 & 4.52 & 1.48 & 2.48 & 2.84 & 3.79 \\
EnhancePPG & \textbf{3.35} & 3.15 & \textbf{2.20} & \textbf{4.38} & \textbf{5.64} & \textbf{2.35} & \textbf{1.93} & \textbf{5.16} & 6.38 & 2.87 & \textbf{3.30} & 5.49 & \textbf{1.84} & \textbf{2.43} & \textbf{2.53} & \textbf{3.54} \\ 
\hline
\end{tabular}}
\vspace{-0.3cm}
\end{table*}
All datasets are downsampled to 32 Hz, using input windows of size \( 4 \times 256 \) (equivalent to 8 seconds) with a 2-second overlap. Each window is normalized per channel using the z-score, computed on the pre-training data when pre-training is applied, or on the training data otherwise. For post-processing, we followed the approach in \cite{qppg}, clipping output values when predictions deviate by more than 10\% from the average of the last 10 estimated values.

\subsection{Ablations and Final HR estimation performance}
\begin{figure}[htpb]
    \centering
    \includegraphics[width=0.95\columnwidth]{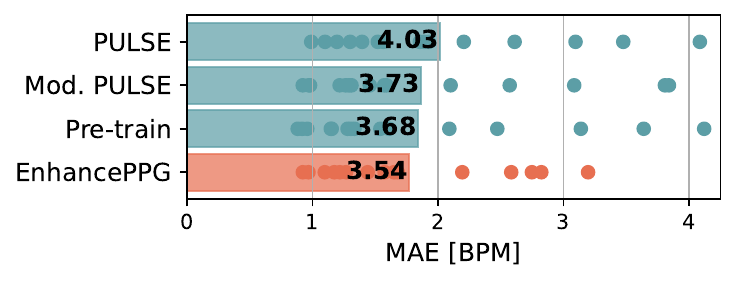}
    \vspace{-0.5cm}
    \caption{MAE results on PPG-DaLia considering architectural modifications (``Mod. PULSE'') and our EnhancePPG (``Pre-train'' and Augmentation).}
    \label{fig:results1}
    \vspace{-0.5cm}
\end{figure}

Figure \ref{fig:results1} illustrates the impact of the different parts of our methodology on the PULSE performance on PPG-DaLiA. Each bar represents the average MAE across 15 subjects, with individual dots indicating the MAE for each patient.

\begin{figure}[htpb]
    \centering
    \includegraphics[width=0.95\columnwidth]{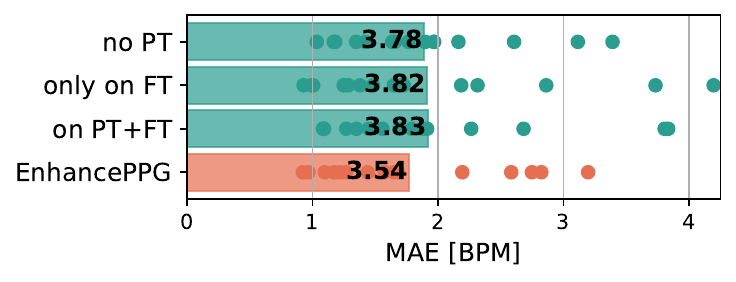}
    \vspace{-0.5cm}
    \caption{MAE results on PPG-DaLiA on best phase to apply DA.}
    \label{fig:results2}
\end{figure}

The original PULSE model achieves an MAE of 4.03 BPM. The \textbf{architectural modifications} described in \ref{sec:pulse_arc} reduce the MAE by 7.44\%, bringing it to 3.73 BPM. This improvement likely stems from multiple factors: the larger kernel size (1, 9) enhances the detection of subtle, short-term variations in PPG and accelerometer signals, while dilated convolutions with dilation 2 may miss critical local fluctuations. Additionally, standard convolutions reduce the over-smoothing effect of dilation and improve noise handling.

Following the architectural modifications, we evaluate our EnhancePPG approach starting with the \textbf{self-supervised pre-training} step described in \ref{sec:pretraining}, where we restructure the model into an autoencoder-like architecture. Using the WESAD dataset and our West Attica dataset for pre-training, we observe a reduction in MAE to 3.68 BPM. While this improvement is modest, it underscores the potential of self-supervised learning, which could deliver even greater performance gains with access to more pre-training data.

\looseness=-1
To verify this intuition and further improve our models' predictive performance, we applied the second step of our approach, i.e. the \textbf{data augmentation} setup described in \ref{sec:aug}, expanding the pre-training dataset by $11\times$. With this setup, referred to as EnhancePPG,  we achieved a MAE of 3.54 BPM, representing a 12.16\% reduction compared to the original PULSE model and a 5.09\% compared to the modified PULSE. This improvement is due to the larger pre-training dataset and the augmentation techniques that expand the BPM range, helping to reduce overfitting and enhance model robustness. We apply augmentation solely on the pre-training, without altering HR labels to prevent generalization issues in supervised training. While \cite{augmentPPG} achieved optimal results by limiting augmentation to specific \textit{extreme} patients, our approach applies it to all patients, resulting in lower overall MAE and enhanced model generalization.
This point is further emphasized in Figure \ref{fig:results2}, where we compare the applications of DA on different training steps. Applying augmentation to the fine-tuning dataset (PPG-DaLiA) in all other configurations, i.e., only during fine-tuning without altering pre-training data (referred to as ``only on FT''), both on pre-training and fine-tuning data (``on PT+FT'') or only to the non-pre-trained model (``no PT''), consistently leads to performance degradation.
Our complete methodology reduces MAE for most patients, with only one exceeding 6 BPM. Table \ref{tab:results_soa} provides detailed per-patient results.

\subsection{Comparison with the state-of-the-art}
Table \ref{tab:results_soa} presents a comparative analysis of recent state-of-the-art methods for HR prediction on the PPG-DaLiA dataset. The KID-PPG result \cite{kid-ppg} is marked with an $^{*}$ symbol due to its use of MA removal, which excludes part of the testing data, making a direct comparison with our method less accurate. Our approach sets a new state-of-the-art, achieving the lowest average MAE, outperforming the previous best, the PULSE model. On a per-subject level, our method reduces the MAE for 11 patients while maintaining comparable performance for the remaining 4. Notably, the MAE for two critical patients (S5 and S8) drops by 18.9\% and 36.8\%, respectively.

Our results show that augmenting pre-training data improves performance without adding complexity or modifying HR labels. This approach maintains consistency between training and test data, unlike \cite{augmentPPG} where modifying HR labels during augmentation led to poorer results.

\subsection{Deployment results}
\begin{table}
\caption{Deployment results on the STM32 NUCLEO-H743ZI2
 }
 \centering
\label{tab:deployment}
\resizebox{0.95\columnwidth}{!}{
\begin{tabular}{c|cccc}
\hline
\textbf{Model} & \textbf{Params.} & \textbf{MACs} & \textbf{Latency} & \textbf{MAE} \\ \hline
PULSE & 230 K & 26.1 M & 413.3 ms & 4.03 BPM \\
Modified PULSE & 386 K & 46.7 M & 431.6 ms & 3.73 BPM \\
EnhancePPG & 386 K & 46.7 M & 431.6 ms & 3.54 BPM \\ \hline
\end{tabular}}
\end{table} 
Table \ref{tab:deployment} outlines the deployment results of our models on the ST32 NUCLEO-H743ZI2, equipped with an Arm Cortex-M7 MCU clocked at 480 MHz. The neural networks were deployed using the ST Edge AI Developer Cloud with Core version 1.0.0. 
The modified PULSE architecture presents an increase in MAC operations and parameters of 78.9\% and 67.8\%, respectively, while maintaining comparable latency, only 4.42\% higher than the original model, given the higher memory regularity of non-dilated convolutions. 
When applied to the modified PULSE, our EnhancePPG further reduces MAE to 3.54 BPM without significantly impacting latency. These significant performance gains come without compromising deployability, making the approach ideal for real-world wearables with similar MCU configurations.

Note that our approach is indeed orthogonal compared to the DL model chosen and can be identically applied to other networks or similar time-series analysis tasks.
\section{Conclusions}
This paper introduced EnhancePPG, an approach to improve heart rate estimation with self-supervised pre-training and data augmentation techniques. These refinements led to a 12.2\% reduction in MAE on the PPG-DaLiA dataset compared to the best SoA model, reducing the error to 3.54 BPM while increasing the overall latency by only 4.42\%. Notably, our approach significantly improved performance for critical patients, making it more applicable for real-world use in wearable health monitoring.

\section*{Acknowledgment}
This publication is part of the project PNRR-NGEU which has received funding from the MUR – DM 117/2023.

\bibliographystyle{IEEEtran}
\bibliography{bibliography}

\begin{thebibliography}{10}
\providecommand{\url}[1]{#1}
\csname url@samestyle\endcsname
\providecommand{\newblock}{\relax}
\providecommand{\bibinfo}[2]{#2}
\providecommand{\BIBentrySTDinterwordspacing}{\spaceskip=0pt\relax}
\providecommand{\BIBentryALTinterwordstretchfactor}{4}
\providecommand{\BIBentryALTinterwordspacing}{\spaceskip=\fontdimen2\font plus
\BIBentryALTinterwordstretchfactor\fontdimen3\font minus \fontdimen4\font\relax}
\providecommand{\BIBforeignlanguage}[2]{{%
\expandafter\ifx\csname l@#1\endcsname\relax
\typeout{** WARNING: IEEEtran.bst: No hyphenation pattern has been}%
\typeout{** loaded for the language `#1'. Using the pattern for}%
\typeout{** the default language instead.}%
\else
\language=\csname l@#1\endcsname
\fi
#2}}
\providecommand{\BIBdecl}{\relax}
\BIBdecl

\bibitem{iot1}
A.~S. Yeole \emph{et~al.}, ``Use of internet of things (iot) in healthcare: A survey,'' in \emph{Proceedings of the ACM Symposium on Women in Research 2016}, 2016.

\bibitem{iot2}
S.~D. Mamdiwar \emph{et~al.}, ``Recent advances on iot-assisted wearable sensor systems for healthcare monitoring,'' \emph{Biosensors}, vol.~11, 2021.

\bibitem{iot3}
R.~Singh \emph{et~al.}, ``Chapter 13 - heart rate monitoring system using internet of things,'' in \emph{IoT-Based Data Analytics for the Healthcare Industry}, 2021.

\bibitem{health-random}
J.~F. Thayer \emph{et~al.}, ``A meta-analysis of heart rate variability and neuroimaging studies: implications for heart rate variability as a marker of stress and health,'' \emph{Neuroscience \& Biobehavioral Reviews}, vol.~36, 2012.

\bibitem{applewatch}
\BIBentryALTinterwordspacing
``Apple, apple watch series.'' [Online]. Available: \url{https://www.apple.com/lae/watch/}
\BIBentrySTDinterwordspacing

\bibitem{fitbit}
\BIBentryALTinterwordspacing
``Fitbit. fitbit charge 4.'' [Online]. Available: \url{https://www.fitbit.com/global/us/products/trackers/charge4}
\BIBentrySTDinterwordspacing

\bibitem{ppg-review}
D.~Biswas \emph{et~al.}, ``Heart rate estimation from wrist-worn photoplethysmography: A review,'' \emph{IEEE Sensors Journal}, vol.~19, 2019.

\bibitem{robustPPG}
N.~Huang \emph{et~al.}, ``Robust ppg-based ambulatory heart rate tracking algorithm,'' in \emph{2020 42nd Annual International Conference of the IEEE Engineering in Medicine \& Biology Society (EMBC)}, 2020.

\bibitem{robustPPG2}
S.~Nabavi \emph{et~al.}, ``A robust fusion method for motion artifacts reduction in photoplethysmography signal,'' \emph{IEEE Transactions on Instrumentation and Measurement}, vol.~69, 2020.

\bibitem{ppg_dalia}
A.~Reiss \emph{et~al.}, ``Deep ppg: Large-scale heart rate estimation with convolutional neural networks,'' \emph{Sensors}, vol.~19, no.~14, 2019.

\bibitem{qppg}
A.~Burrello \emph{et~al.}, ``Q-ppg: Energy-efficient ppg-based heart rate monitoring on wearable devices,'' \emph{IEEE Transactions on Biomedical Circuits and Systems}, vol.~15, 2021.

\bibitem{embedPPG}
------, ``Embedding temporal convolutional networks for energy-efficient ppg-based heart rate monitoring,'' \emph{ACM Trans. Comput. Healthcare}, vol.~3, 2022.

\bibitem{augmentPPG}
------, ``Improving ppg-based heart-rate monitoring with synthetically generated data,'' in \emph{2022 IEEE Biomedical Circuits and Systems Conference (BioCAS)}, 2022.

\bibitem{kid-ppg}
\BIBentryALTinterwordspacing
C.~Kechris \emph{et~al.}, ``Kid-ppg: Knowledge informed deep learning for extracting heart rate from a smartwatch,'' 2024. [Online]. Available: \url{https://arxiv.org/abs/2405.09559}
\BIBentrySTDinterwordspacing

\bibitem{pulse}
P.~Kasnesis \emph{et~al.}, ``Feature-level cross-attentional ppg and motion signal fusion for heart rate estimation,'' in \emph{2023 IEEE 47th Annual Computers, Software, and Applications Conference (COMPSAC)}, 2023.

\bibitem{u-net}
O.~Ronneberger \emph{et~al.}, ``U-net: Convolutional networks for biomedical image segmentation,'' in \emph{Medical Image Computing and Computer-Assisted Intervention -- MICCAI 2015}, 2015.

\bibitem{wesad}
P.~Schmidt \emph{et~al.}, ``Introducing wesad, a multimodal dataset for wearable stress and affect detection,'' in \emph{Proceedings of the 20th ACM International Conference on Multimodal Interaction}, 2018.

\bibitem{biseizure}
\BIBentryALTinterwordspacing
L.~Benfenati \emph{et~al.}, ``Biseizure: Bert-inspired seizure data representation to improve epilepsy monitoring,'' 2024. [Online]. Available: \url{https://arxiv.org/abs/2406.19189}
\BIBentrySTDinterwordspacing

\bibitem{loso1}
A.~Reiss \emph{et~al.}, ``Ppg-based heart rate estimation with time-frequency spectra: A deep learning approach,'' in \emph{Proceedings of the 2018 ACM International Joint Conference and 2018 International Symposium on Pervasive and Ubiquitous Computing and Wearable Computers}, 2018.

\bibitem{loso2}
Y.~Zhang and ohters, ``Ppg-based heart rate estimation with efficient sensor sampling and learning models,'' in \emph{2022 IEEE 24th Int Conf on High Performance Computing \& Communications; 8th Int Conf on Data Science \& Systems; 20th Int Conf on Smart City; 8th Int Conf on Dependability in Sensor, Cloud \& Big Data Systems \& Application (HPCC/DSS/SmartCity/DependSys)}, 2022.

\bibitem{empatica}
\BIBentryALTinterwordspacing
``Empatica. e4 wristband 2014.'' [Online]. Available: \url{https://www.empatica.com/en-eu/research/e4/}
\BIBentrySTDinterwordspacing

\end{thebibliography}

\end{document}